# Topology optimization based on moving deformable components – A new computational framework

Xu Guo[1], Weisheng Zhang and Wenliang Zhong

*State Key Laboratory of Structural Analysis for Industrial Equipment,*

*Department of Engineering Mechanics,*

*Dalian University of Technology, Dalian, 116023, P.R. China*

**Abstract**

In the present work, a new computational framework for structural topology optimization based on the concept of moving deformable components is proposed. Compared with the traditional pixel or node point-based solution framework, the proposed solution paradigm can incorporate more geometry and mechanical information into topology optimization directly and therefore render the solution process more flexible. It also has the great potential to reduce the computational burden associated with topology optimization substantially. Some representative examples are presented to illustrate the effectiveness of the proposed approach.

---

[1]Corresponding author.    E-mail:  guoxu@dlut.edu.cn    Tel: +86-411-84707807

# 1. Introduction

Structural topology optimization, which aims at placing available material within a prescribed design domain appropriately in order to achieve optimized structural performances, has received considerable research attention since the pioneering work of Bendsoe and Kikuchi [1]. Many approaches have been proposed for structural topology optimization and it now has been extended to a wide range of physical disciplines such as acoustics, electromagnetics and optics. We refer the readers to [2-5] and the references therein for a state-of-the-art review of topology optimization.

From geometry representation point of view, most of the existing topology optimization methods are developed within the pixel or node point-based solution framework. For example, in the well-established artificial density with penalization approach [6-8], the design domain is first discretized into finite elements (pixels) with reasonable resolution, then mathematical programming or optimality criteria-based algorithms are applied to find the element-wise black-and-white (i.e., 0 or 1 in each pixel) density distribution, which represents the topology of the structure, see Fig. 1 for reference. Although remarkable achievements have been made by this approach, there are still some challenging issues need further explorations. Firstly, it is worth noting that the pixel-based geometry/topology representations not consistent with that in modern Computer-Aided-Design (CAD) modeling systems, where the geometries/topologies of structures are often described by geometric primitives such as points, line segments or Bezier curves and the corresponding Boolean operations between them (See Fig. 2 for reference). Therefore topology optimization cannot be conducted on CAD platform directly. Secondly, since no geometry information is embedded in the pixel-based topology optimization approaches explicitly, it is difficult to give a precise control of the structural feature sizes (i.e., minimum/maximum length scale, minimum curvature), which is usually very important from manufacturing considerations. Finally, since the element-wise material distribution is utilized to represent the structural topology, the computational efforts involved in pixel-based topology optimization approaches are relatively large especially when three-dimensional problems are considered. For example, if the design domain is a $1 \times 1 \times 1$ cubic as shown in Fig. 3 and discretized by 100 elements along each direction (note that this is only a relatively low resolution), the number of the design variables will increase to one million, which is far beyond the solution capacity of existing mathematical optimization algorithms.

For the node point-based topology optimization approaches, level-set method [9, 10] is a representative one. In the level set method, it is also needed to discretize the design domain into finite elements to calculate the structural responses, the values of the level set function at the node points, however, are often used as topological design variables, see Fig. 4 for reference. The structural boundary (topology) can be identified by extracting the zero contour of the level set function. Although geometry information such as the normal outward vector and curvature of the boundary can be calculated from the level set function, level set method basically suffers from the same disadvantage of the variable density method since its implicit geometry representation is also quite different from the explicit one adopted in CAD modeling systems. Furthermore, node point-based level set method also cannot escape from the curse of dimensionality as mentioned before.

With the primary aim of establishing a direct link between structural topology optimization and CAD modeling systems and therefore conducting topology optimization in a more geometry explicit and flexible way, in the present work, a new computational framework for structural topology optimization based on the concept of moving deformable components is proposed. Compared with the traditional pixel or node point-based solution framework, the proposed solution paradigm can incorporate more geometry and mechanical information into topology optimization directly and therefore render the solution process more flexible. Furthermore, it also has the great potential to reduce the computational burden associated with topology optimization substantially.

The rest of the paper is organized as follows. In Section 2, the basic idea of the proposed approach is explained in detail. The corresponding numerical solution aspects are discussed in Section 3. A comparison of the new solution framework with the existing ones is made in Section 4. In Section 5, some representative examples are presented to illustrate the effectiveness of the proposed approach. Finally, some concluding remarks are provided in Section 6.

## 2. Moving deformable components-based topology optimization framework

In this section, the basic idea of the proposed topology optimization approach will be explained in detail.

### 2.1 Deformable component-the primary building block of topology optimization

Traditional topology optimization approaches, both pixel-based and node point-based ones, are basically based on the framework of ground structure. In this framework, one first fills the entire design domain with material and then deletes the unnecessary parts from it gradually. This is achieved by updating the element densities in variable density approaches while the evolution of structural boundaries in level set approaches, respectively. In the present work, we propose a different solution framework for topology optimization, where deformable components are intended to use as primary building blocks.

To illustrate the basic idea, let us consider the topology optimization of a short-beam shown in Fig. 5a, which is designed to transmit a vertical load to the clamped support with minimum structural compliance (maximum stiffness) under available volume constraint. It is well known that a manufacturable optimal solution takes the form shown in Fig. 5b. It can be observed from this figure that the optimal structure is constituted by eight "structural components". Hereinafter, a structural component represents an amount of material occupied a specific volume in the design domain. In fact, any structure with any type of topology can also be decomposed into a finite number of components. The above observation inspires us that these "structural components" may be used as the basic building blocks of topology optimization as shown schematically in Fig. 6. The optimal structural topology can be obtained by determining the geometry characteristic parameters, such as the shape, length, thickness and orientation as well as the layout (connectivity) of these components, through optimality conditions. With use of this idea, a computational framework for topology optimization of continuum structures, which is quite different from the previous ones and capable of incorporating more geometry information into problem formulation, can be established. It is worth noting that in the proposed solution framework, the components are allowed to be overlapped with each other. It is just through this overlapping mechanism, the layout of the structure are changed and optimized. In the present framework, redundant components can be "disappeared" (in the sense that it has no influence on structural responses) through being overlapped by another component. This is quite different from the traditional approaches where unnecessary materials are deleted from the design domain through some degeneration mechanisms to achieve structural topology changes. In some sense, the proposed approach can be viewed as an *adaptive ground structure approach* for topology optimization. This is actually the main source where its advantages come from. We will come back to this point in the following discussions.

**2.2 Geometry description of a structural component**

In this subsection, we shall discuss how to describe the geometry of a structural component using explicit parameters. As a primary attempt to develop the present new computational framework, a relatively simple form of building component is introduced in the present work. However, we will also discuss how to deal with the more general situations where structural components with more complex geometries are involved at the end of this subsection. For the sake of simplicity, only two-dimensional (2D) case is considered here. Extensions to three-dimensional (3D) case will be discussed in a separate work.

If the topology of a structure is the main concern of a structural design and optimization problem, structural components with rectangular shapes can serve as the basic building blocks of topology optimization especially when the number of involved components is relatively large. As shown in Fig. 7, even a small number of components can represent various fairly complicated structural topologies. Mathematically, the region $\Omega$ occupied by a rectangular component centered at $(x_0, y_0)$ with length $L$, thickness $t$ and inclined angle $\theta$ (measured from the horizontal axis anti-clockwisely) can be described by the following level set function (see Fig. 8 for reference)

$$\begin{cases} \phi(x) > 0, & \text{if } x \in \Omega, \\ \phi(x) = 0, & \text{if } x \in \partial\Omega, \\ \phi(x) < 0, & \text{if } x \in D \setminus \Omega, \end{cases} \quad (2.1)$$

where

$$\phi(x,y) = \left(\frac{\cos\theta \cdot (x - x_0) + \sin\theta \cdot (y - y_0)}{L/2}\right)^n + \left(\frac{-\sin\theta \cdot (x - x_0) + \cos\theta \cdot (y - y_0)}{t/2}\right)^n - 1$$

and $n$ is a relatively large even integer number. The structural component can move, dilate/shrink and rotate in the design domain by changing the values of $x_0, y_0, L, t$ and $\theta$. If there are totally $n$ structural components in the design domain, the structural topology can be described implicitly as

$$\begin{cases} \boldsymbol{\phi}^s(\boldsymbol{x}) > 0, & \text{if } \boldsymbol{x} \in \Omega^s, \\ \boldsymbol{\phi}^s(\boldsymbol{x}) = 0, & \text{if } \boldsymbol{x} \in \partial\Omega^s, \\ \boldsymbol{\phi}^s(\boldsymbol{x}) < 0, & \text{if } \boldsymbol{x} \in D\backslash\Omega^s, \end{cases} \quad (2.2)$$

where $\boldsymbol{\phi}^s = (\phi_1, \ldots, \phi_n)^\top$ with $\phi_i, i = 1, \ldots, n$ denoting the topology description function of the $i$-th component and $\Omega^s$ is the region occupied by the solid structural components.

## 2.3 Topology optimization based on moving deformable components: problem formulation

Based on the discussion above, we propose the following formulation for moving deformable component-based topology optimization:

$$\text{Find } \boldsymbol{d} = (\boldsymbol{d}_1, \ldots, \boldsymbol{d}_{nc})^\top$$

$$\text{Minimize } I = I(\boldsymbol{d})$$

s.t.

$$g_i(\boldsymbol{d}) \leq 0, \quad i = 1, \ldots, m,$$

$$\boldsymbol{d} \subset \mathcal{U}_d, \quad (2.3)$$

where the symbol $nc$ denotes the total number of components involved in the optimization problem. The symbol $\boldsymbol{d} = (\boldsymbol{d}_1, \ldots, \boldsymbol{d}_{nc})^\top$ represents the vector of design variables with $\boldsymbol{d}_i = (x_{0i}, y_{0i}, L_i, t_i, \theta_i)^\top, i = 1, \ldots, nc$. $\mathcal{U}_d$ is the admissible sets that $\boldsymbol{d}$ belongs to. In Eq. (2.3), $g_i, i = 1, \ldots, m$ are the considered constraint functionals.

If the considered topology optimization is to minimize the compliance of the structure under available volume constraint, the problem formulation can be specified as

$$\text{Find } \boldsymbol{d} = (\boldsymbol{d}_1, \ldots, \boldsymbol{d}_{nc})^\top$$

$$\text{Minimize } I = I(\boldsymbol{d}, \boldsymbol{u})$$

s.t.

$$\int_{\Omega^s = \cup_{i=1}^{nc} \Omega_i} \mathbb{E}(\boldsymbol{x}) : \boldsymbol{\varepsilon}(\boldsymbol{u}) : \boldsymbol{\varepsilon}(\boldsymbol{v}) \mathrm{dV} = \int_{\Omega^s = \cup_{i=1}^{nc} \Omega_i} \boldsymbol{f}(\boldsymbol{x}) \cdot \boldsymbol{v} \mathrm{dV} + \int_{\Gamma_t} \boldsymbol{t} \cdot \boldsymbol{v} \mathrm{dS}, \quad \forall \boldsymbol{v} \in \mathcal{U}_{\text{ad}},$$

$$V(\boldsymbol{d}) \leq \bar{V},$$

$$d \subset \mathcal{U}_d,$$
$$u = \bar{u}, \quad \text{on } \Gamma_u. \tag{2.4}$$

In Eq. (2.4), $\Omega_i$, $i = 1, \ldots, nc$ denote the region occupied by the $i$-th component, respectively. $u$ and $v$ are the displacement field and the corresponding test function defined on $\Omega^s = \bigcup_{i=1}^{nc} \Omega_i$ with $\mathcal{U}_{ad} = \{v | v \in \mathbf{H}^1(D), v = 0 \text{ on } \Gamma_u\}$. $f$ and $t$ denote the body force density in $\Omega_i$, $i = 1, \ldots, nc$ and the surface traction on Neumann boundary $\Gamma_t$ of $\Omega^s$, respectively. $\bar{u}$ is the prescribed displacement on Dirichlet boundary $\Gamma_u$. The symbol $\varepsilon$ denotes the second order linear strain tensor. In Eq. (2.4), $\mathbb{E} = E/(1+v)[\mathbb{I} + v/(1-2v)\delta \otimes \delta]$ ($\mathbb{I}$ and $\delta$ denote the fourth and second order identity tensor, respectively) is the fourth order isotropic elasticity tensor with $E$ and $v$ denoting the corresponding Young's modulus and Poisson's ratio, respectively. To be more specific, $\mathbb{E}(x) = \mathbb{E}^i, f(x) = f^i$, where $\mathbb{E}^i$ and $f^i$ are the corresponding values associated with the $i$-th component, respectively. Furthermore, the symbol $\bar{V}$ denotes the upper bound of the available materials volume.

It is worth noting that if the above topology optimization is solved with use of Eulerian description and fixed finite element mesh on a prescribed design domain $D$, the corresponding problem formulation can be expressed in terms of $\phi^s$ as

$$\text{Find} \quad d = (d_1, \ldots, d_{nc})^\top$$
$$\text{Minimize} \quad I = I(d, u)$$

s.t.

$$\int_D H(\phi^s) \mathbb{E} : \varepsilon(u) : \varepsilon(v) \, dV = \int_D H(\phi^s) f \cdot v \, dV + \int_{\Gamma_t} t \cdot v \, dS, \quad \forall v \in \mathcal{U}_{ad},$$

$$\int_D H(\phi^s) \, dV \leq \bar{V},$$

$$d \subset \mathcal{U}_d,$$

$$u = \bar{u}, \quad \text{on } \Gamma_u, \tag{2.5}$$

where, for the sake of simplicity, the assumption that all structural components are made from the same type material is adopted. For the case where multi-materials are considered, one can resort to the so-called "color level set" representation, which will not be touched in the present study for the limitation of space.

*Remark 1.* A natural problem associated with the proposed approach is how many structural components should be included in the problem formulation? Of course, it can be expected that the more components are included, the better the optimal solutions will be. In fact, theoretical analysis indicates that optimal solutions of topology optimization problems always contain microstructures constituted by infinite numbers of "bars" with infinitesimal thickness. If the manufacturability and robustness (e.g., refrain from buckling under compressive forces) of the design is taken into consideration, however, only structures with finite number of structural components are of practical use in engineering applications. Due to this fact, it seems reasonable to include only limited number of structural components in the problem formulation. It is also worth noting that in the problem formulation, it is not necessary to eliminate the redundant structural components completely to achieve topology degenerations since the structural components can change their lengths and positions freely. This means that a long bar can be constituted by several short bars. Furthermore, as shown in Fig. 9, a redundant component can also be disappeared by hiding itself into another larger component. This feature is very helpful to circumvent the singularity phenomena which are often related to the degeneration of materials in traditional topology optimization approaches [11, 12]. In fact, allowing the overlapping of structural components is the key point for the success of the proposed approach.

*Remark 2.* Although in the present study, only structural components with rectangular shapes are considered, the proposed computational framework does has the potential to account for the case where structural components with curved boundaries are involved. This is due to the fact that on the one hand, any curved structural components can be approximated with controllable accuracy through a number of rectangular structural components both geometrically and mechanically (see Fig. 10 for reference). On the other hand, we can also introduce appropriate implicit geometry design variables to optimize the positions, shapes and the layout of a set of curved structural components directly. For example, as illustrated in Fig. 11, this can be achieved by optimizing the shape of the skeleton (see [13] and [14] for its definition and applications) of a structural component, which can be described by the well-established Non-Uniform Rational B-Spline (NURBS), and the thicknesses of the component at some specific interpolation points (i.e., $t_i$). We will not pursue the details on this aspect here and intend to report the corresponding results in a separate work.

## 3. Numerical solution aspects

In this section, we shall discuss the numerical solution aspects of the proposed component-based topology optimization framework.

### 3.1 Finite element analysis

In the proposed optimization framework, the background finite element mesh is fixed and the boundary of a structural component is described *implicitly* by an *explicit* level set function. This treatment is very flexible to deal with the possible overlapping of the structural components, which is the key mechanism to achieve topology changes in the proposed optimization framework. In view of this, the XFEM analysis based on the level set description of structural geometries is adopted for structural analysis [15]. With use of this approach, re-meshing is only needed in the vicinity of structural boundaries in order to enhance the accuracy of displacement/stress computations (see Fig. 12 for reference). Furthermore, we also need weak material ($E = 1.0e - 06$) to mimic voids in the design domain, which is indispensible to establish the interactions between different components before the final optimized structure is obtained.

### 3.2 Sensitivity analysis

For a general optimization problem where the objective functional can be written as a volume integral such that

$$I = \int_D H(\boldsymbol{\phi}^s) F(\boldsymbol{u}) \mathrm{dV}, \tag{2.6}$$

where $\boldsymbol{\phi}^s = (\phi_1, \dots, \phi_n)^\top$ is the level set function of the entire structure with $n$ denoting the total number of structural components in the design domain. Under this circumstance, the variation of $I$ with respect to the variation of individual $\phi_i$ can be calculated as

$$\Delta I = \int_D \delta_i(\boldsymbol{\phi}^s) f(\boldsymbol{u}, \boldsymbol{w}) \Delta \phi_i \mathrm{d}V, \tag{2.7}$$

where $\boldsymbol{u}$ is the primary displacement field and $\boldsymbol{w}$ is the adjoint displacement field, which can be determined by solving a corresponding adjoint boundary value problem. In Eq. (2.7), $f$ is a function of $\boldsymbol{u}$ as well as $\boldsymbol{w}$ and $\delta_i(\boldsymbol{\phi}^s) = \min(\delta(\phi_i), \delta(\boldsymbol{\phi}^s))$.

For the considered optimization problem (i.e., structural compliance minimization under volume constraint and therefore $I = \int_D H(\boldsymbol{\phi}^s) \mathbb{E}: \boldsymbol{\varepsilon}(\boldsymbol{u}): \boldsymbol{\varepsilon}(\boldsymbol{u}) \mathrm{d}V$ ), we have $\boldsymbol{w} = \boldsymbol{u}$ and $f(\boldsymbol{u}, \boldsymbol{w}) = f(\boldsymbol{u}, \boldsymbol{u}) = -2\mathbb{E}: \boldsymbol{\varepsilon}(\boldsymbol{u}): \boldsymbol{\varepsilon}(\boldsymbol{u})$ since it is a self-adjoint problem. It is also straightforward to obtain that

$$\Delta \phi_i = f_{i1} \Delta x_{0i} + f_{i2} \Delta y_{0i} + f_{i3} \Delta L_i + f_{i4} \Delta t_i + f_{i5} \Delta p_i, \tag{2.8}$$

with

$$f_{i1} = n \left( \frac{q_i(x - x_{0i}) + p_i(y - y_{0i})}{(L_i/2)} \right)^{n-1} \frac{-q_i}{(L_i/2)} + n \left( \frac{-p_i(x - x_{0i}) + q_i(y - y_{0i})}{(t_i/2)} \right)^{n-1} \frac{p_i}{(t_i/2)}, \tag{2.9a}$$

$$f_{i2} = n \left( \frac{q_i(x - x_{0i}) + p_i(y - y_{0i})}{(L_i/2)} \right)^{n-1} \frac{-p_i}{(L_i/2)} + n \left( \frac{-p_i(x - x_{0i}) + q_i(y - y_{0i})}{(t_i/2)} \right)^{n-1} \frac{-q_i}{(t_i/2)}, \tag{2.9b}$$

$$f_{i3} = \frac{1}{2} n \left( \frac{q_i(x - x_{0i}) + p_i(y - y_{0i})}{(L_i/2)} \right)^{n-1} \frac{q_i(x - x_{0i}) + p_i(y - y_{0i})}{-(L_i/2)^2}, \tag{2.9c}$$

$$f_{i4} = \frac{1}{2} n \left( \frac{-p_i(x - x_{0i}) + q_i(y - y_{0i})}{(t_i/2)} \right)^{n-1} \frac{-p_i(x - x_{0i}) + q_i(y - y_{0i})}{-(t_i/2)^2}, \tag{2.9d}$$

$$f_{i5} = n \left( \frac{q_i(x - x_{0i}) + p_i(y - y_{0i})}{(L_i/2)} \right)^{n-1} \left( \frac{-p_i(x - x_{0i})/q_i + (y - y_{0i})}{(L_i/2)} \right)$$
$$+ n \left( \frac{-p_i(x - x_{0i}) + q_i(y - y_{0i})}{(t_i/2)} \right)^{n-1} \left( \frac{-(x - x_{0i}) - p_i(y - y_{0i})/q_i}{(t_i/2)} \right), \tag{2.9e}$$

where $p_i = \sin\theta_i$ and $q_i = \cos\theta_i = \sqrt{1 - p_i^2}$, respectively.

In summary, we have

$$\frac{\partial I}{\partial x_{0i}} = -2 \int_D \delta_i(\boldsymbol{\phi}^s) \mathbb{E}: \boldsymbol{\varepsilon}(\boldsymbol{u}): \boldsymbol{\varepsilon}(\boldsymbol{u}) f_{i1} \mathrm{d}V, \tag{2.10a}$$

$$\frac{\partial I}{\partial y_{0i}} = -2 \int_D \delta_i(\boldsymbol{\phi}^s) \mathbb{E}: \boldsymbol{\varepsilon}(\boldsymbol{u}): \boldsymbol{\varepsilon}(\boldsymbol{u}) f_{i2} \mathrm{d}V, \tag{2.10b}$$

$$\frac{\partial I}{\partial L_i} = -2 \int_D \delta_i(\boldsymbol{\phi}^s) \mathbb{E}: \boldsymbol{\varepsilon}(\boldsymbol{u}): \boldsymbol{\varepsilon}(\boldsymbol{u}) f_{i3} \mathrm{d}V, \tag{2.10c}$$

$$\frac{\partial I}{\partial t_i} = -2 \int_D \delta_i(\boldsymbol{\phi}^s) \mathbb{E} : \boldsymbol{\varepsilon}(\boldsymbol{u}) : \boldsymbol{\varepsilon}(\boldsymbol{u}) f_{i4} \mathrm{dV}, \qquad (2.10d)$$

$$\frac{\partial I}{\partial p_i} = -2 \int_D \delta_i(\boldsymbol{\phi}^s) \mathbb{E} : \boldsymbol{\varepsilon}(\boldsymbol{u}) : \boldsymbol{\varepsilon}(\boldsymbol{u}) f_{i5} \mathrm{dV}, \qquad (2.10e)$$

## 4. Merits of the proposed topology optimization framework

As mentioned in the introduction, topology optimization has undergone tremendous development during the last three decades. Many approaches have been proposed since the pioneering work of Bendsoe and Kikuchi (1988). Then a problem arises naturally: why another new topology optimization framework? The following are some discussions on this point.

From the authors' point of view, compared with existing topology optimization approaches, the proposed one has the following distinctive features:

(1) As discussed in the previous sections, it is obvious that the proposed method has a natural link with the CAD modeling systems since the geometries of the basic building blocks of optimization, the structural components, are described *explicitly* by parameterized surfaces/curves, which are the basic operable objects in computer graphics. This is quite different from the traditional pixel-based variable density method and the *implicit* surface-based level set method. In some sense, the proposed way of geometry description is quite consistent with the modern language of differential geometry where a complex manifold, roughly speaking, can be represented by a series of overlapping parametrizable patches. This feature makes it possible that the proposed approach can not only being integrated with CAD systems seamlessly but also give an *explicit* and *local* control of the structural features in a natural way. Furthermore, as shown clearly in the previous discussions, in the proposed method, the optimization model is totally independent of the analysis model. This is very helpful to circumvent the numerical problems such as checkerboard patterns, mesh-dependency of optimal solutions, which often appear in traditional topology optimization frameworks. Another advantage is that since the geometry of the structural components are described explicitly in the present computational framework, possible uncertainties of the components shapes usually arising from manufacture errors can also be dealt with in a relatively direct way compared to existing approaches [16-18]. Finally, it is also worth noting that the complete independence of the optimization model and analysis model may also give us more freedom to develop non-FEM based topology optimization methods, which are very needed for multi-physics applications.

(2) The proposed method has the capability to integrate shape, size and topology optimization or even structural type optimization, where the appropriate types of structural components (i.e., beam, shell, plane membrane) are sought for, in a unified framework. In the proposed method, the shape and size of individual structural components can be optimized by changing their geometry description parameters (e.g., coordinates of the interpolation points) while the optimal structural topology can be obtained by varying the connectivity of the structural components. The later can be achieved through appropriate positioning and overlapping of the components. As for the optimal selection of the type of structural components, we can discretize the individual structural components using different types of finite elements (e.g., beam element, shell element and plane membrane element) and impose the relevant geometrical constraints in order to make the corresponding structural mechanics theory applicable (e.g., the characteristic length to thickness ratio is greater than 10 for a beam component). With use of optimization algorithms, we can determine the optimal layout of these structural elements and therefore achieve the (structural element) type optimization. The above treatment may also help eliminate the possible inconsistency between the optimization model and the analysis model (e.g., modeling a slender beam with a small number of plane membrane elements), which often exists in traditional topology optimization frameworks and open a new avenue in practical application of topology optimization.

(3) The proposed method has great potential to share the merits of both Lagrangian and Eulerian topology optimization approaches. This is due to the fact that on the one hand, we have crisp description of the structural boundaries, which provides a natural advantage of dealing with boundary-dependent loads or complex boundary conditions by constructing body fitted meshes especially in multi-physics settings. On the other hand, since the structural components are allowed to overlap with each other in the proposed solution framework, the intrinsic flexibility of Eulerian description for describing the change of structural topology has also been inherited successfully.

(4) The proposed method is a pure black-and–white one since in fact only layout optimization of *solid* structural components is utilized to achieve the variation of structural topology. In view of this, some intrinsic difficulties associated with the traditional approaches (e.g., the variable density approach) such as the suppression of gray elements and the construction of rational interpolation schemes (especially for multi-physics problem) can be totally eliminated. Furthermore manipulating pure black-and–white designs can also help accelerate the convergence rate since no grey elements, which are the main sources preventing the optimization algorithms from rapid converging [5], exist during the entire course of optimization.

(5) The proposed method has great potential to reduce the computational efforts associated with topology optimization. In the proposed method, geometry description parameters of the structural components are adopted as design variables. As a consequence, the number of design variables may be quite smaller than that involved in traditional topology optimization approaches. For example, for the short beam problem discussed in Section 2, if we include 20 structural components with rectangular shape in the initial design, the total number of design variables is 100, which includes 40 coordinate variables ($x_{0i}, y_{0i}, i = 1, ...,20$), 40 thickness and length variables ($t_i, L_i, i = 1, ...,20$) and 20 inclined angle variables ($\theta_i, i = 1, ...,20$), respectively. It is also worth noting that this number is totally independent of the finite element resolution used for structural analysis. On the contrary, for the variable density or level set method, the total number of design variable, is more than 3000 even for a relatively low 80X40 mesh resolution! There is no doubt that this reduction of the number of design variables will be even more remarkable for 3D problems. This reduction of design space is very important to enhance the efficiency of topology optimization and circumvent the curse of dimensionality since the computation complexity of an optimization problem increases almost linearly with respect to the number of design variables. Besides, the relatively small size of design space also makes it possible that global optimization methods can be utilized to find the optimal designs with global optimality and surrogate models can be constructed to further alleviate the computational efforts associated with the structural responses analysis. Furthermore, since geometry description of each component is totally independent, the proposed method has the intrinsic parallelism, which can be further utilized to enhance the computational efficiency.

We will further explore the above advantages of the proposed computational framework in a series of subsequent research works.

## 5. Numerical examples

In this section, the proposed moving deformable components-based topology optimization approach is applied to several numerical examples for demonstration of its effectiveness. Since the main purpose of the present study is to examine the numerical performance of the proposed algorithm and not to design real life structures, the material, load and geometry data are all chosen as dimensionless. Only 2D plane stress problems with unit thickness are considered. The displacement fields are solved approximately with use of uniform bilinear square elements. Furthermore, Method of Moving Asymptotes (MMA) [19] is adopted to solve the optimization problems numerically.

### 5.1 The short beam example

The problem under investigation is plotted in Fig. 13. The displacement is set to zero along the left side of the design domain. First, let us consider the case where a unit vertical load is imposed on the middle point of the right side (Point A in Fig. 13). The dimensions of the initial design domain are $L = 2$ and $W = 1$, respectively. The design domain is discretized by a $100 \times 50$ FEM mesh. The design objective is to minimize the mean compliance of the structure under the available solid material constraint such that $\bar{V} \leq 0.5$.

The initial design shown in Fig. 14 is composed of 16 components which can move, rotate, dilate and shrink during the process of optimization. The corresponding optimal topology is shown in Fig. 15. The value of the objective functional is $I = 69.44$.

This result is almost the same as that obtained by the other methods, for example SIMP and level set methods. But it is worth mentioning that the number of the design variable using the proposed method is only 80. On the contrary, for the variable density or level set method, the total number of design variable, is more than 3000 for the same FEM mesh. It means that the computational cost can be dramatically reduced with use of the proposed method. Fig. 16 shows some intermediate steps of the optimization process. Considering the fact that a little change of the component may lead to a big change of topology, a relatively small step length is adopted. Even under this circumstance, a optimal solution can be achieved within 100 iterations. The numerical results illustrate clearly the flexibility and capability of the proposed method to handle drastic topological changes. Table 1 lists the value of design variables corresponding to the optimal solution. It is worth noting that although the boundaries of the solutions in Fig. 15a and Fig. 16 look like zigzag, it is only a manifestation of the contour extraction algorithm on a coarse mesh. In fact, if we plot the result in a CAD system (also shown in the same figure), smooth boundary can be observed. From Fig. 15b, the layout of the components can also be observed clearly. Note that in the CAD plot, the components whose widths are less than one mesh width have been plotted.

Next, let us consider the case where the unit vertical load is imposed on the right bottom of the design domain (Point B in Fig. 13). The other parameters are the same as previous example. Starting from the same initial design as shown in Fig. 14, we can obtain the optimal design shown in Fig. 17, which is very close to the solution obtained by classical methods. The corresponding value of the objective functional is $I = 78.54$. Some intermediate steps of numerical optimization are shown in Fig.18 and the values of optimal design variables are listed in Table 2, respectively. From Fig. 18, it can be observed that during the course of optimization, the sizes of the components locating at the regions with small strain energy densities (i.e., component 13) will reduce to small values and those components locating at central region of the design domain will merge into a single one gradually. This is quite reasonable from optimization point of view.

## 5.2 The MBB example

This is another well-known benchmark example usually used for examining the numerical performance of a topology optimization approach. The design domain, boundary conditions, geometry data and external load are all shown in Fig. 19. Since the problem under consideration is symmetric in nature, only half of the design domain is taken into account and discretized by a $120 \times 40$ FEM mesh. The initial design shown in Fig. 20 is composed of 24 components which is similar to that in the previous example. As the same in the previous example, the structure is optimized to minimize the mean compliance of the structure under the available solid material constraint $\bar{V} \leq 0.4$.

The optimal topology is shown in Fig. 21 and the corresponding value of the objective functional is $I = 234.10$. This result is very similar to that obtained by traditional methods but obtained with only 120 design variables! This is a significant reduction of design variables compared to traditional methods (4800 design variables for the same FEM mesh). In addition, some intermediate steps of numerical optimization are shown in Fig. 22 and the values of optimal design variables are listed in Table 3, respectively. It can be observed from Fig. 22 that some components will first get close, then overlap and finally merge into a single one during the process of optimization, which is in fact the critical mechanism to allow for the change of structural topology. The numerical results of this example indicate once again that the proposed method dose have the capability to deal with topology optimization problems.

## 6. Concluding remarks

In the present work, a moving deformable component-based theoretical framework for structural topology optimization is suggested. Unlike in the traditional solution frameworks, where topology optimization is achieved by eliminating unnecessary materials from the design domain or evolving the structural boundaries, optimal structural topology is obtained by optimizing the layout of deformable structural components in the proposed approach. To the best of the authors' knowledge, this is a novel idea which has not been explored in literature. One of the advantages of the proposed approach, which may have great potential in engineering applications, is that it can integrate the size, shape and topology optimization in CAD modeling systems seamlessly. Of course, the proposed method is still in the stage of infancy. Although the new solution framework seems attractive from theoretical point of view and the presented examples do have shown its potentials, more work need to be done to explore its efficiency, initial design-dependency, robustness and rate of convergence especially for non-self-adjoint, large scale and multi-physics oriented topology optimization problems. Corresponding research results will be reported elsewhere.

**Acknowledgements**

The financial supports from the National Natural Science Foundation (10925209, 91216201, 11372004), 973 Project of China (2010CB832703), Program for Changjiang Scholars, Innovative Research Team in University (PCSIRT) and 111 Project (B14013) are gratefully acknowledged.

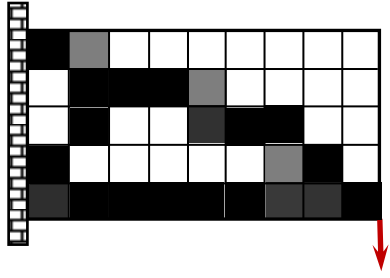

Fig.1 Pixel-based topology optimization

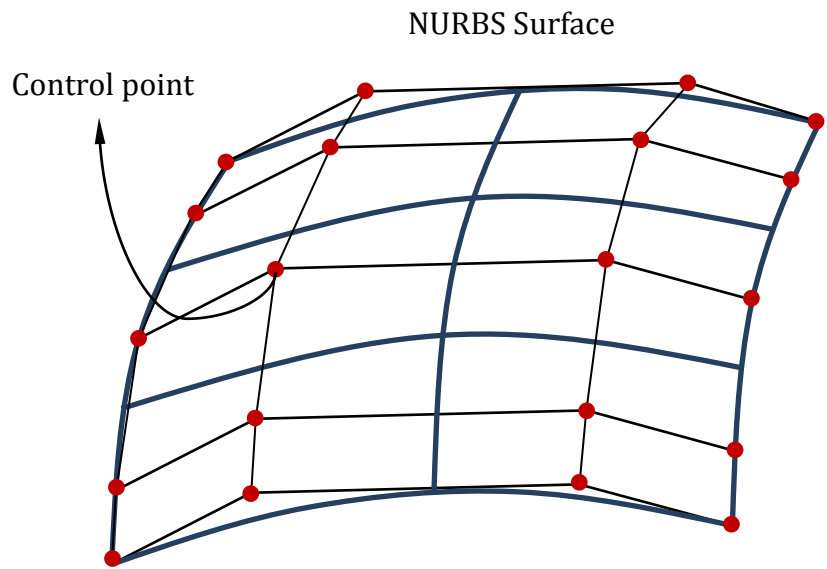

Fig. 2 Geometry and topology representation in CAD system

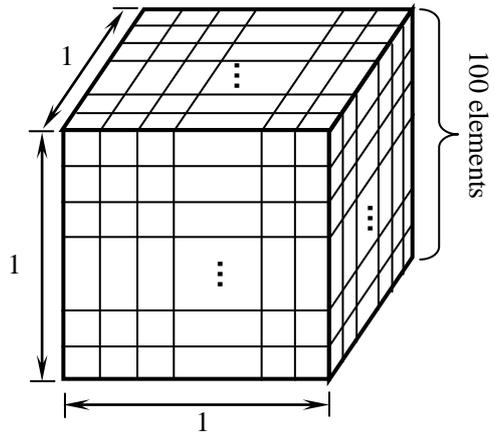

Fig. 3 The curse of dimensionality in topology optimization

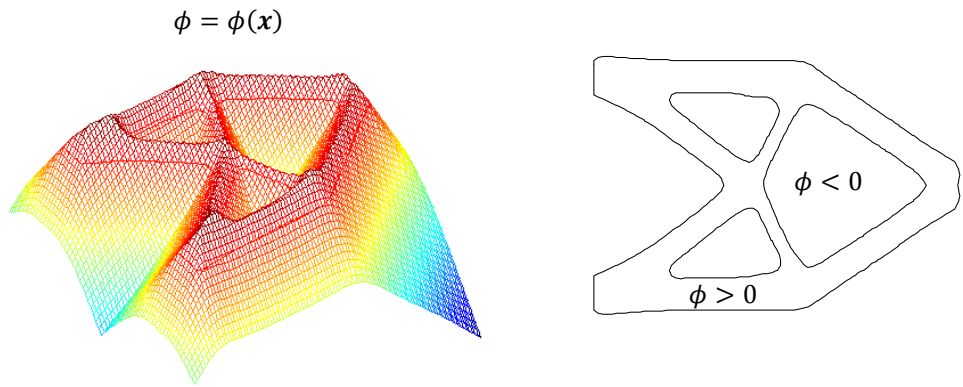

Fig. 4 Node point-based topology optimization

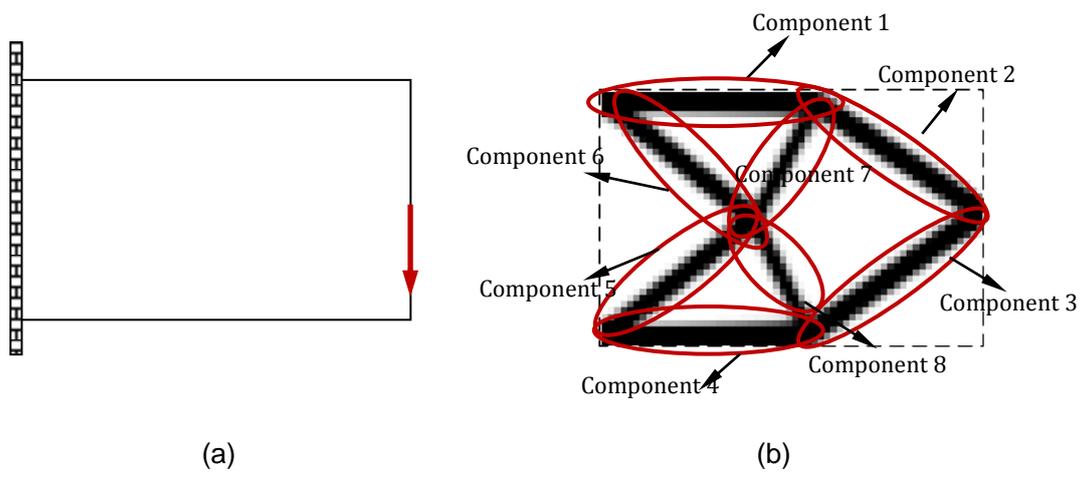

(a)                  (b)

Fig. 5 Structural topology represented by the layout of structural components

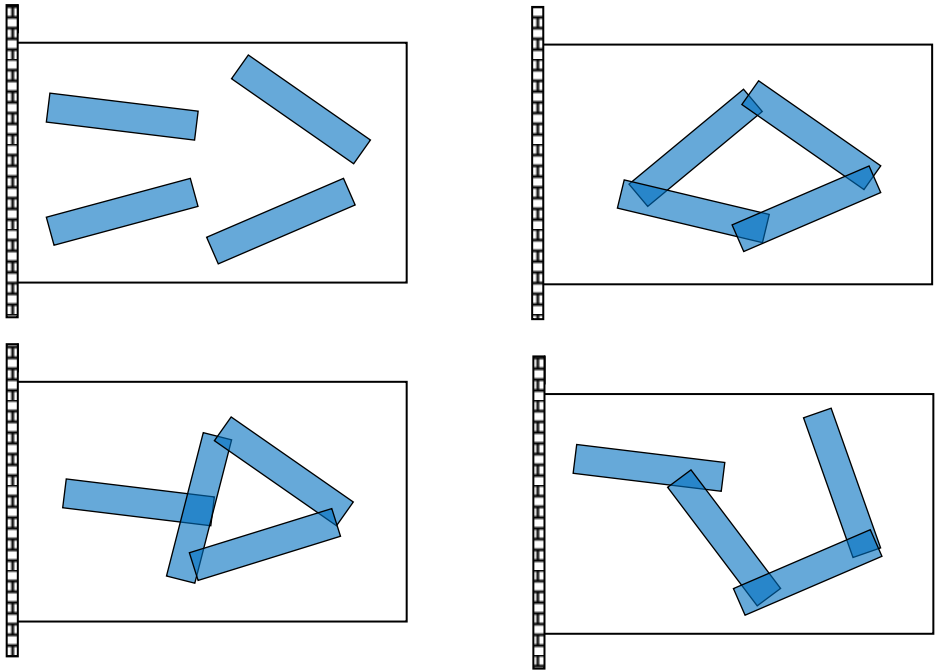

Fig. 6 Structural components as basic building blocks of topology optimization

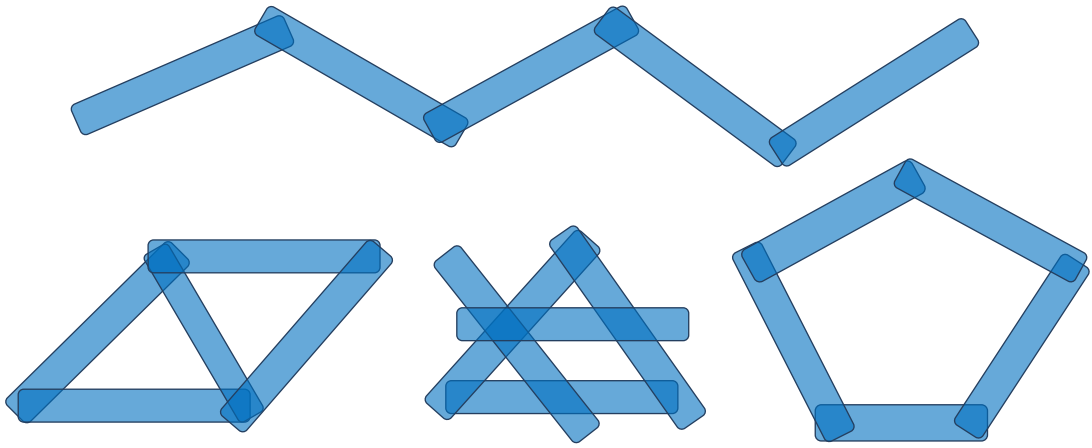

Fig. 7 Simple components and complex structural topologies

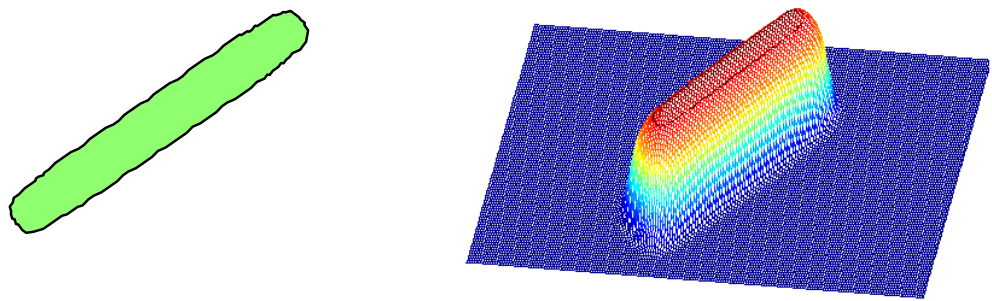

Fig. 8 Rectangular structural component and its level set function

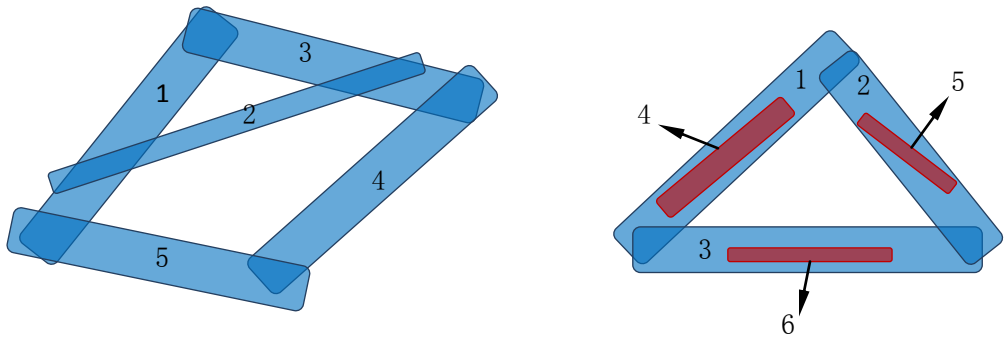

Fig. 9 Topology variation through hiding mechanism of components

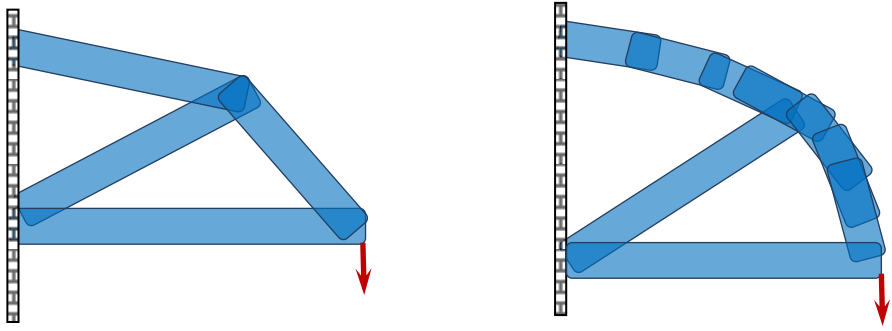

Fig. 10 Approximation of curved structural components with use of straight ones

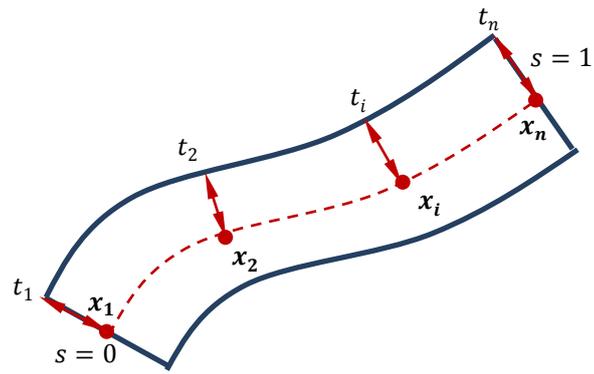

Fig. 11 Skeleton-based topology optimization

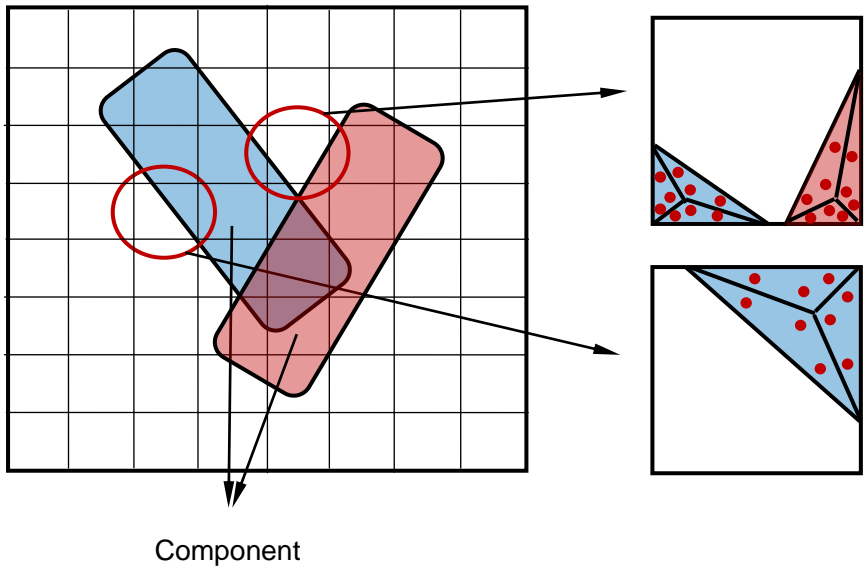

Component

Fig. 12 XFEM analysis based on a fixed mesh

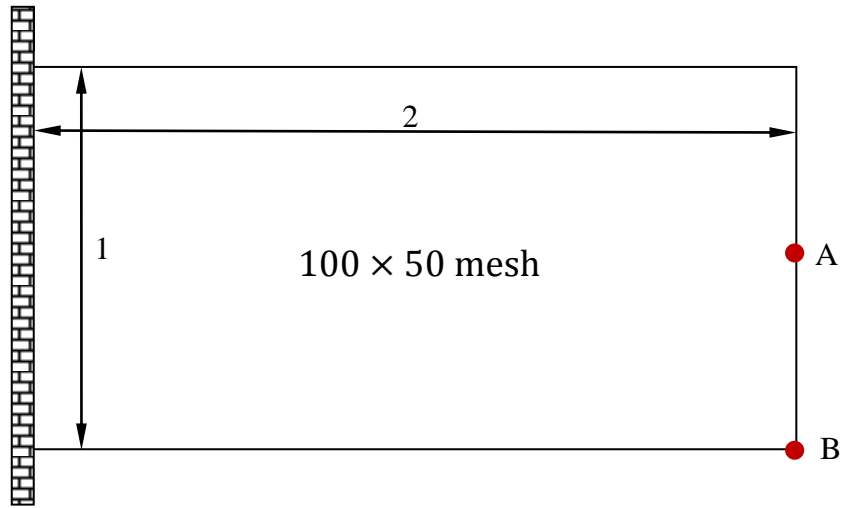

Fig. 13 The short beam example

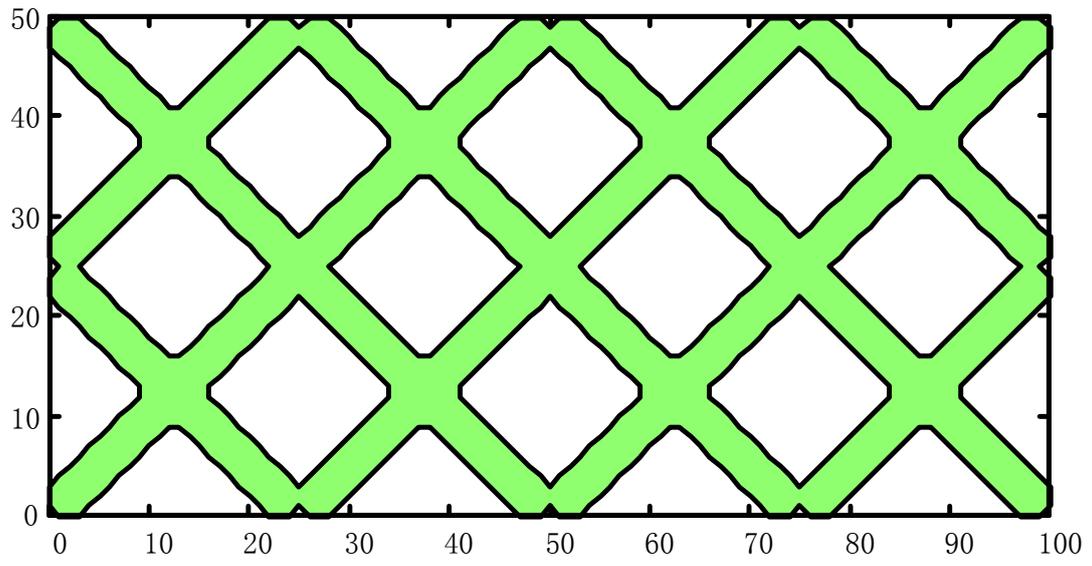

Fig. 14 Initial design for the short beam example

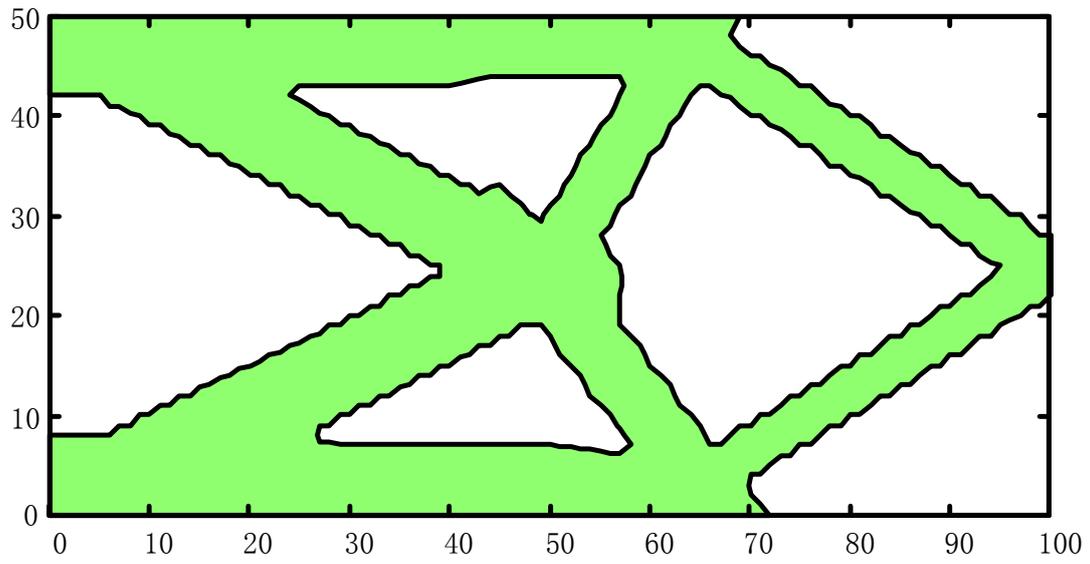

(a) Contour plot

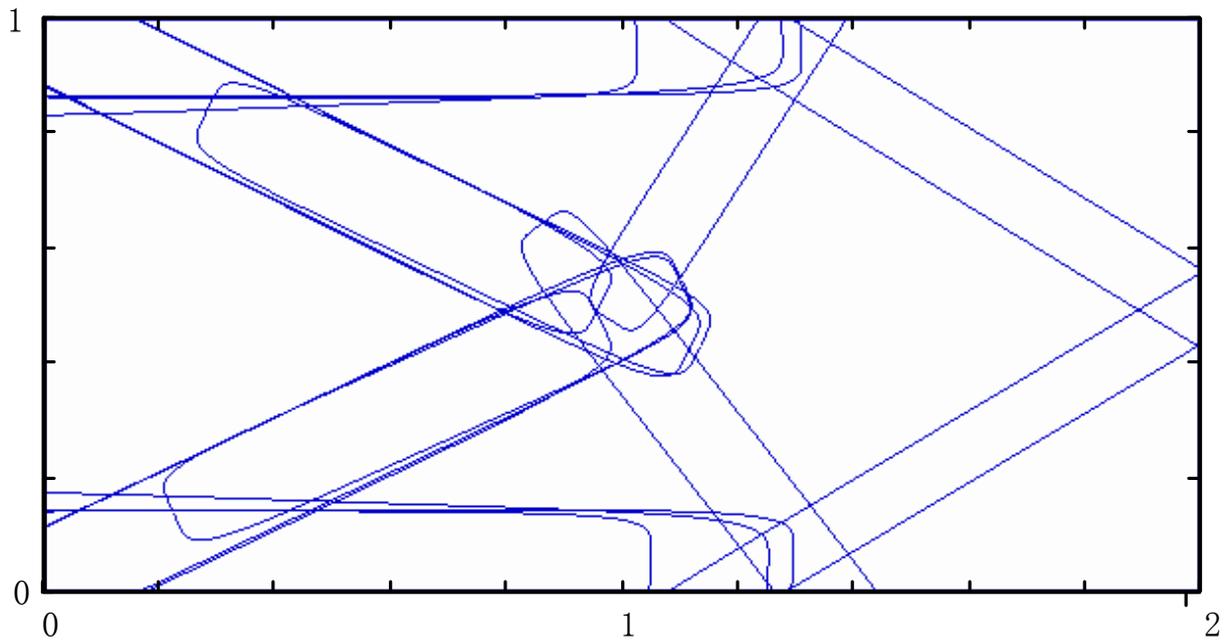

(b) CAD plot

Fig. 15 Optimal topology of the short beam example

(load imposed at Point A)

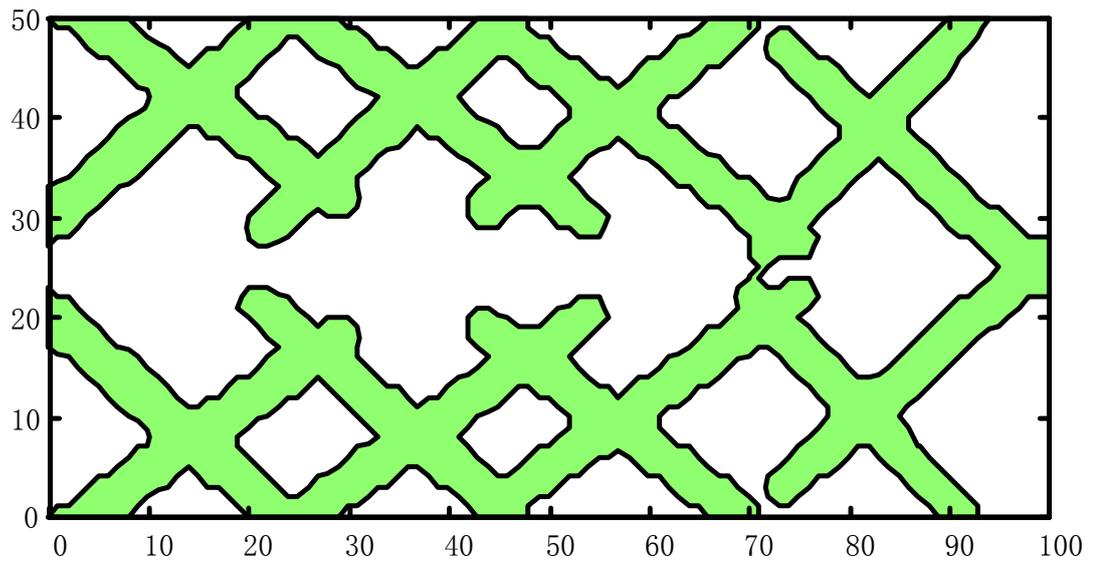

Step 5

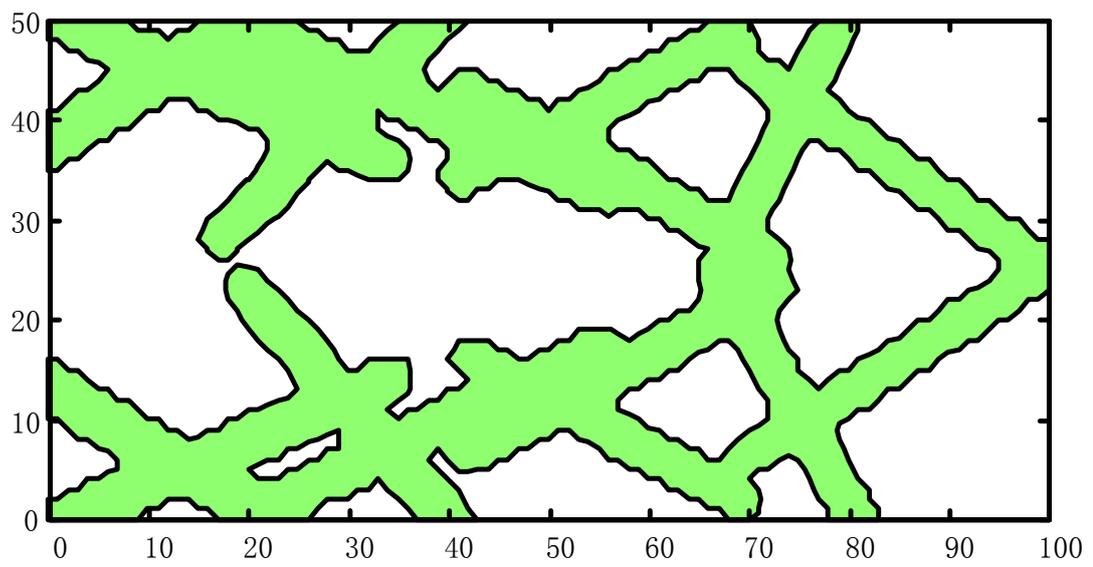

Step 20

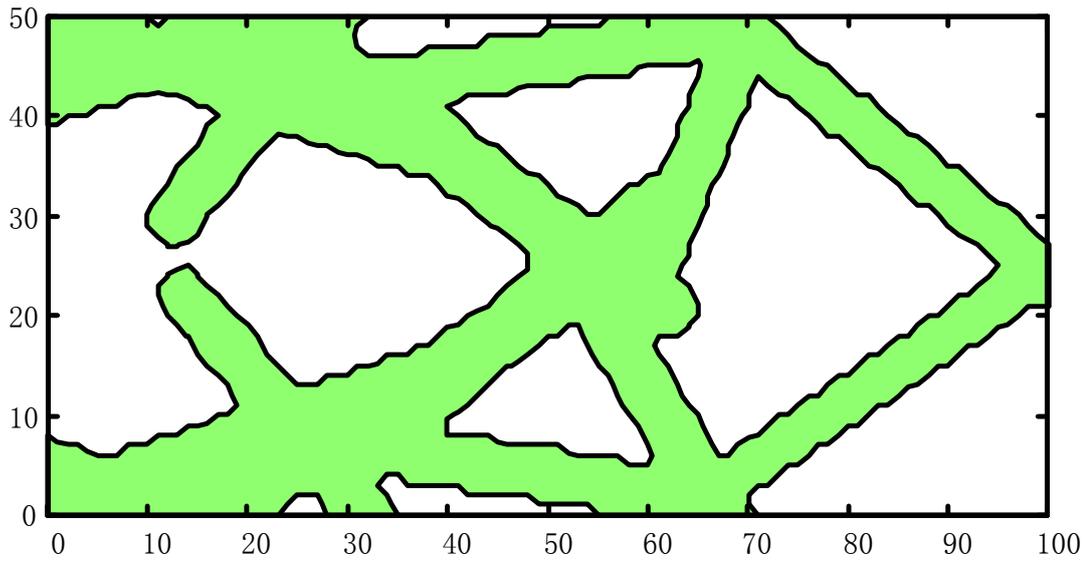

Step 60

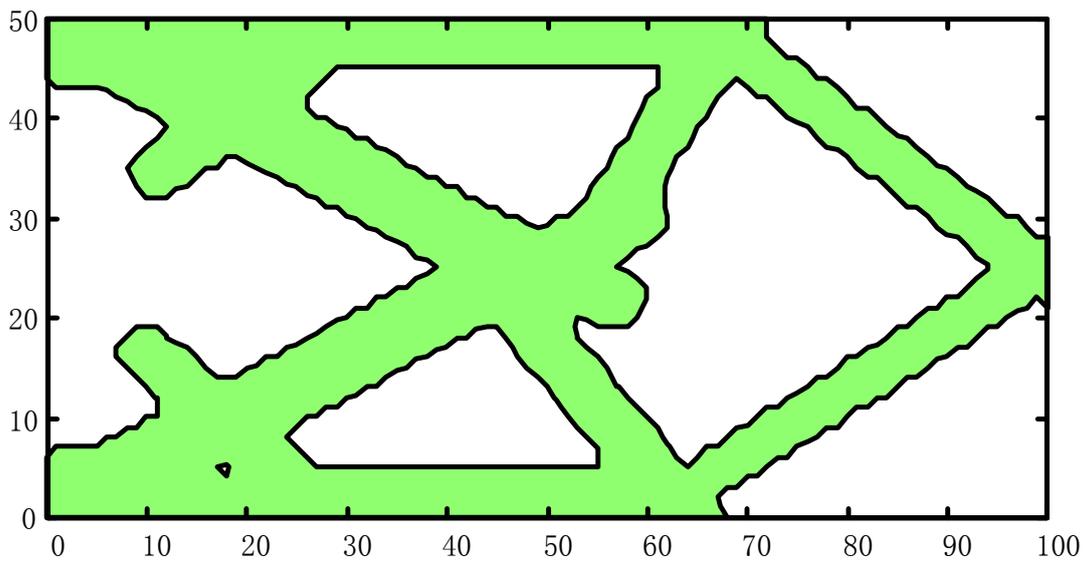

Step 80

Fig. 16 Some intermediate iteration steps of the short beam example

(load imposed at Point A)

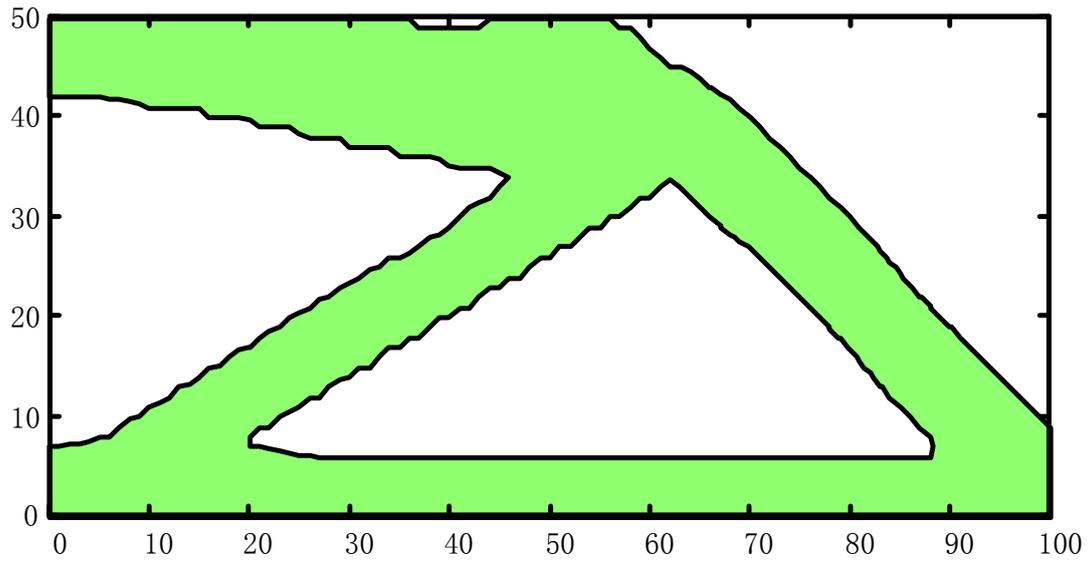

(a) Contour plot

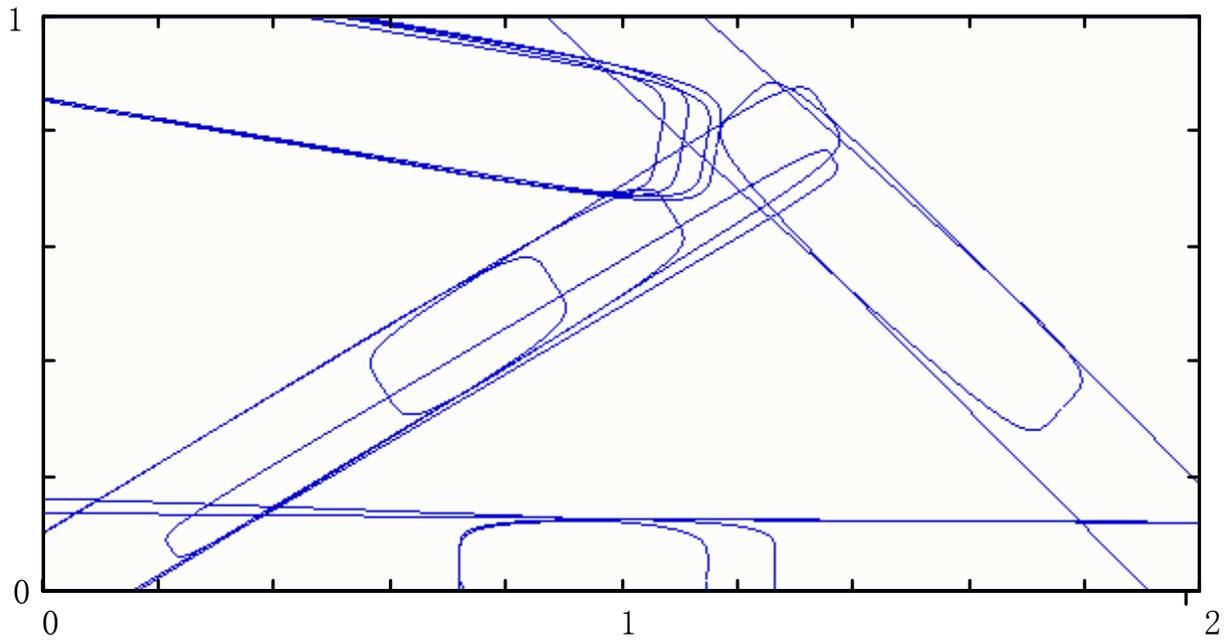

(b) CAD plot

Fig. 17 Optimal topology of the short beam example

(load imposed at Point B)

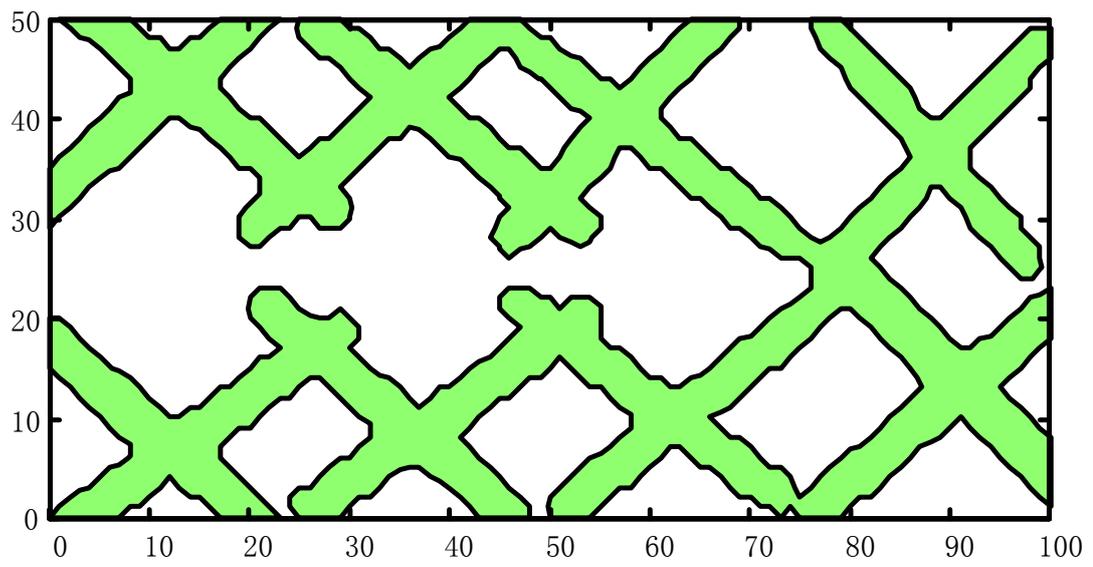

Step 5

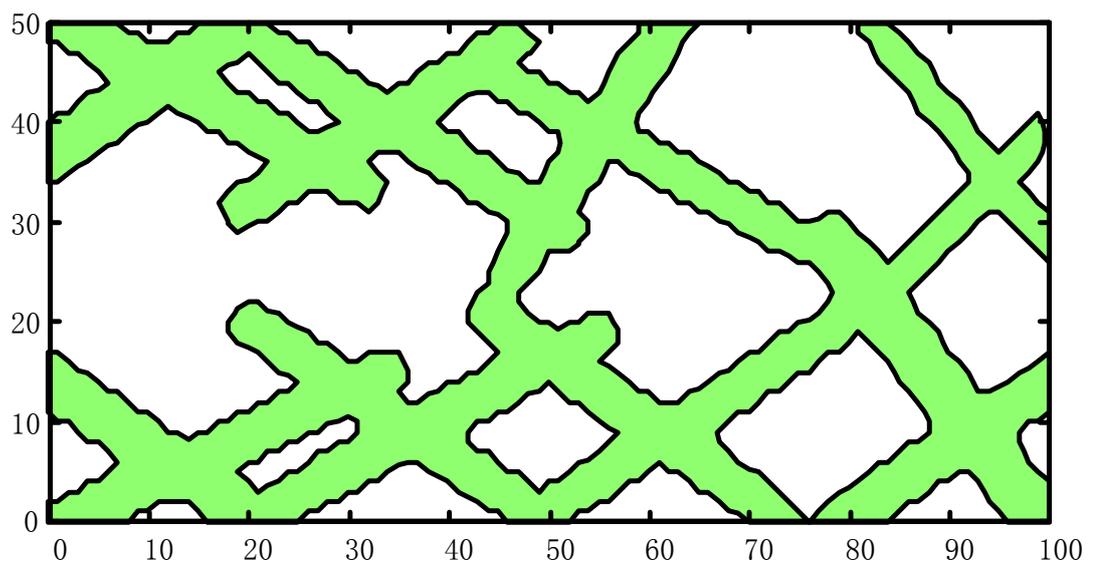

Step 20

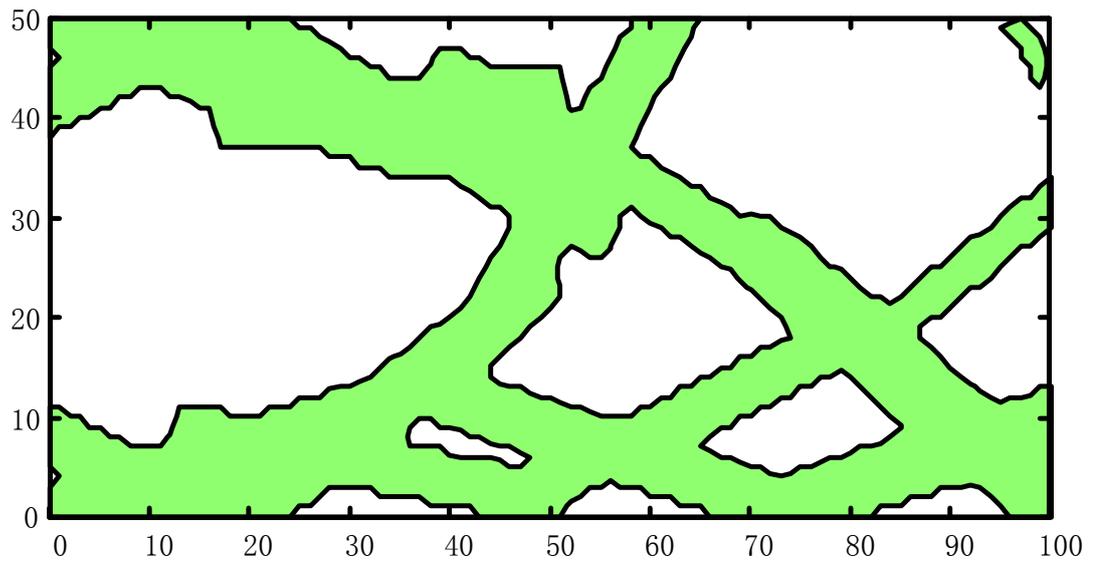

Step 60

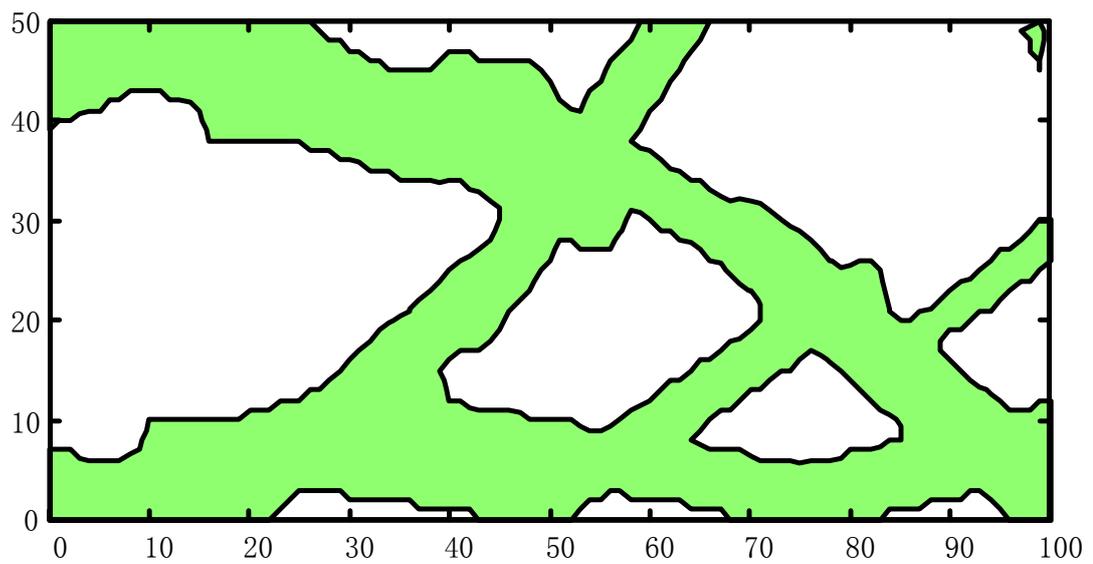

Step 80

Fig. 18 Some intermediate iteration steps of the short beam example
(load imposed at Point B)

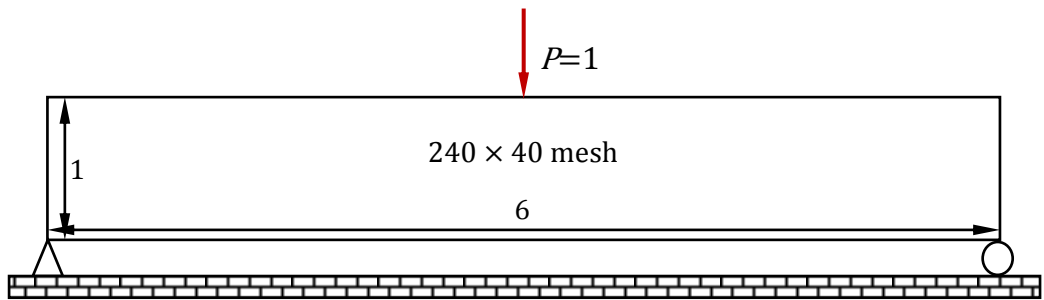

Fig. 19 The MBB example

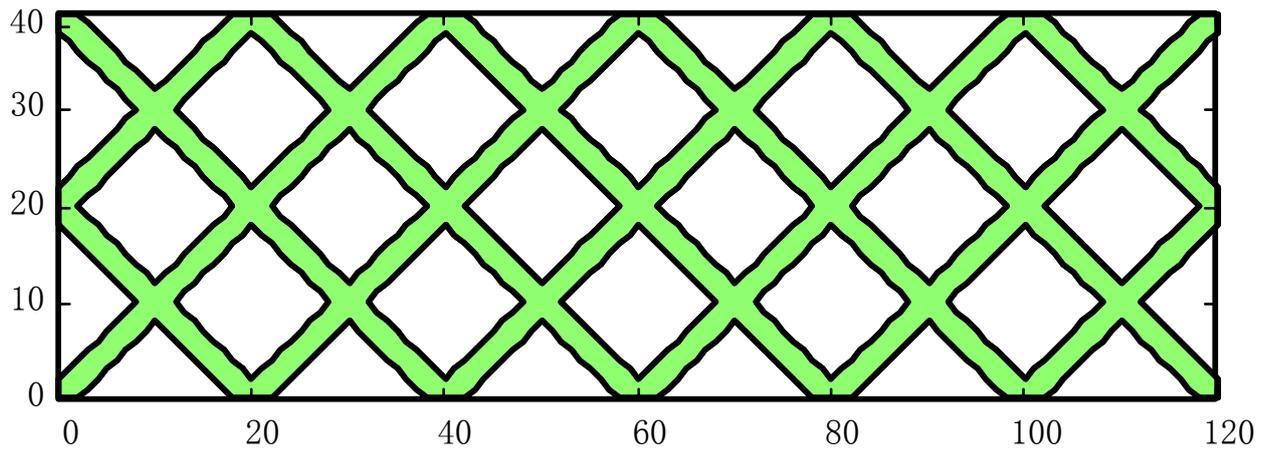

Fig. 20 The initial design of the MBB example

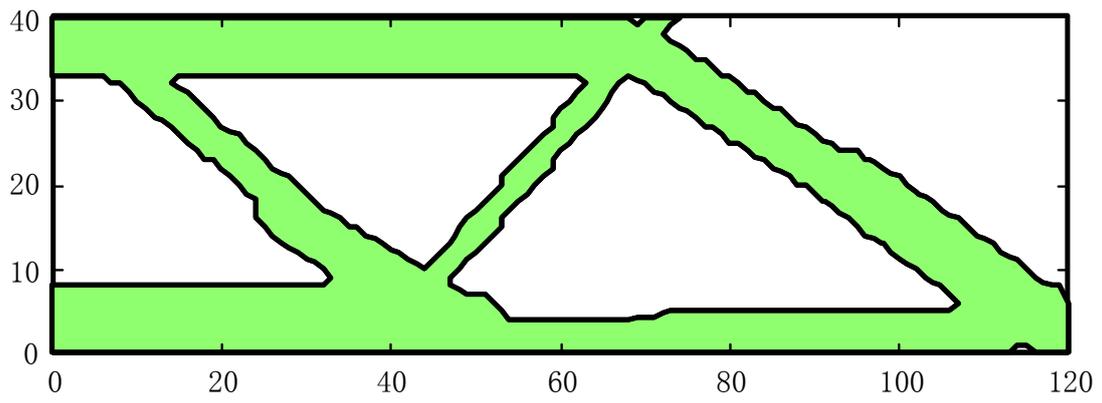

(a) Contour plot

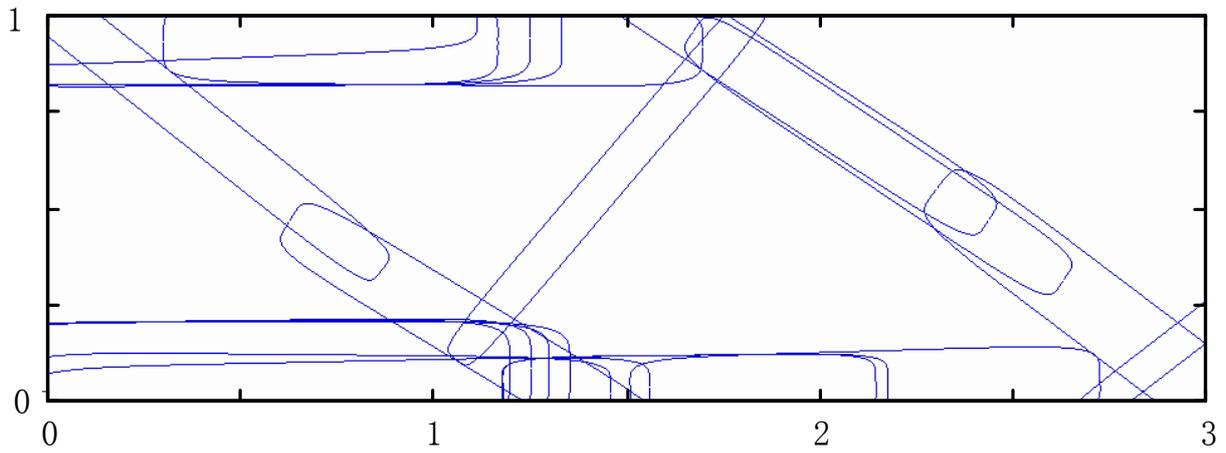

(b) CAD plot

Fig. 21 Optimal topology of the MBB example (half)

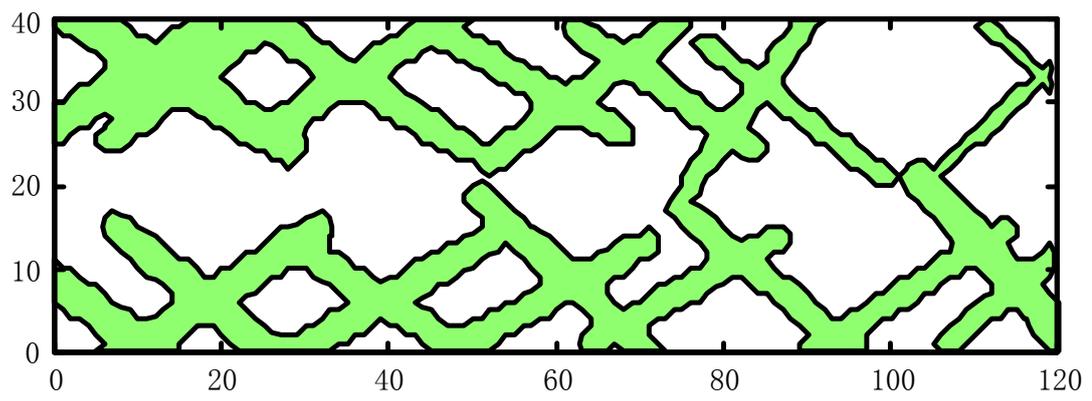

Step 20

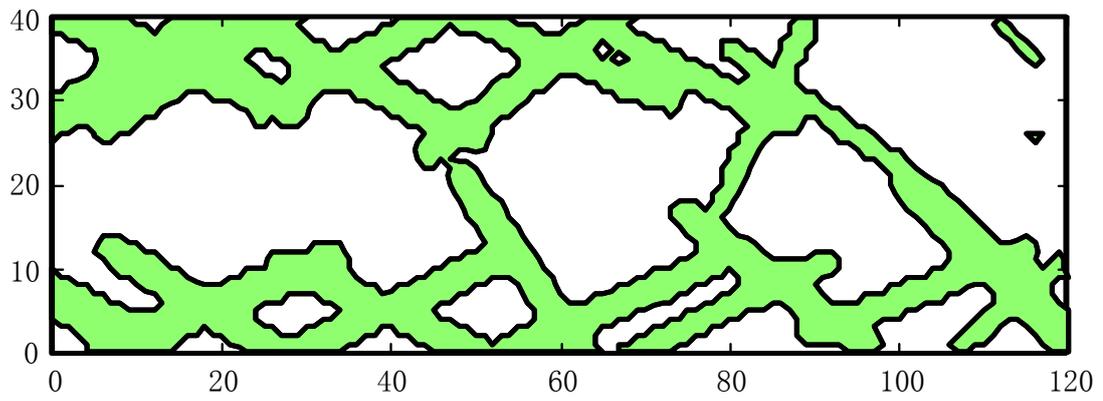

Step 40

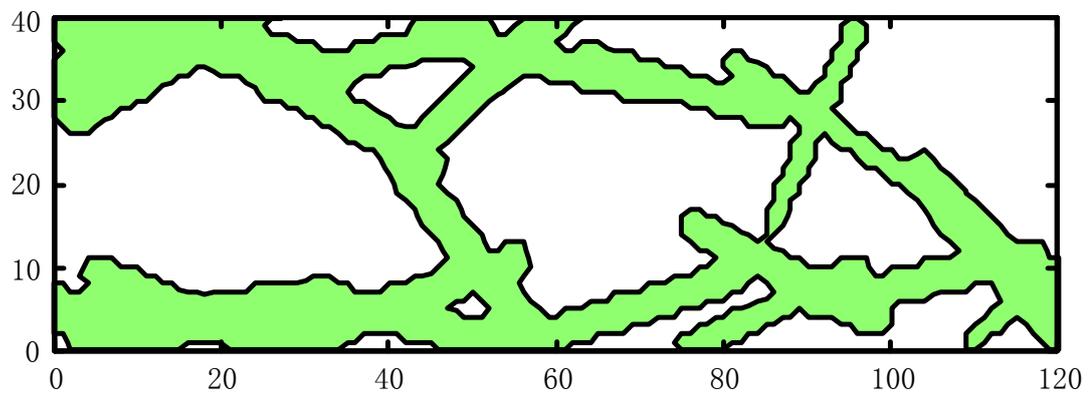

Step 60

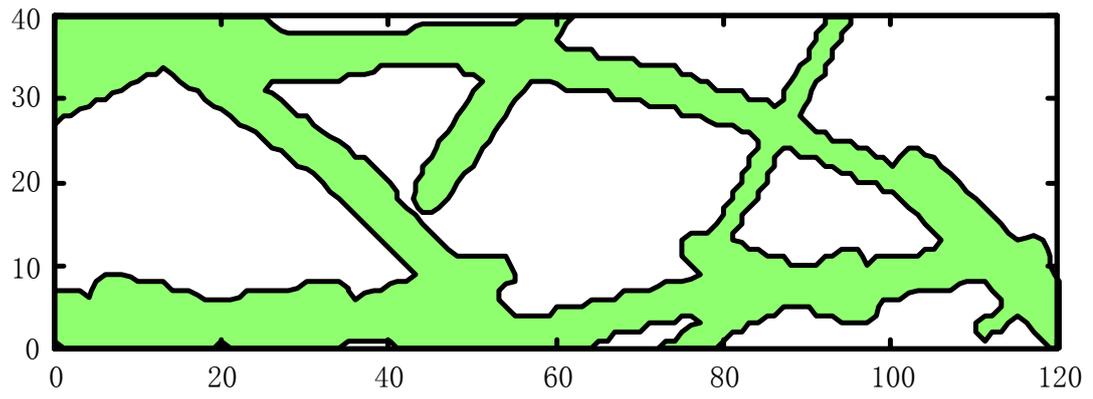

Step 80

Fig. 22 Some intermediate iteration steps of the MBB example (half)

Table 1. Optimal solution of the short beam example (load imposed at Point A)

| Component | $x_0$ | $y_0$ | $L/2$ | $t/2$ | $p$ |
|---|---|---|---|---|---|
| 1 | 0.53 | 0.95 | 0.75 | 0.10 | 0.04 |
| 2 | 0.31 | 0.83 | 0.73 | 0.09 | 0.45 |
| 3 | 0.32 | 0.16 | 0.72 | 0.09 | 0.43 |
| 4 | 0.51 | 0.05 | 0.74 | 0.10 | 0.03 |
| 5 | 0.43 | 0.95 | 0.59 | 0.08 | 0.00 |
| 6 | 0.41 | 0.78 | 0.79 | 0.09 | 0.45 |
| 7 | 0.40 | 0.21 | 0.78 | 0.09 | 0.43 |
| 8 | 0.49 | 0.06 | 0.57 | 0.08 | 0.01 |
| 9 | 0.54 | 0.95 | 0.77 | 0.08 | 0.00 |
| 10 | 0.71 | 0.63 | 0.47 | 0.08 | 0.44 |
| 11 | 0.66 | 0.34 | 0.48 | 0.08 | 0.41 |
| 12 | 0.53 | 0.06 | 0.77 | 0.08 | 0.00 |
| 13 | 1.20 | 0.82 | 0.40 | 0.06 | 0.84 |
| 14 | 1.65 | 0.72 | 0.75 | 0.06 | 0.53 |
| 15 | 1.62 | 0.26 | 0.79 | 0.05 | 0.52 |
| 16 | 1.18 | 0.22 | 0.53 | 0.07 | 0.79 |

Table 2. Optimal solution of the short beam example (load imposed at Point B)

| Component | $x_0$ | $y_0$ | $L/2$ | $t/2$ | $p$ |
|---|---|---|---|---|---|
| 1 | 0.39 | 0.91 | 0.73 | 0.12 | -0.18 |
| 2 | 0.39 | 0.90 | 0.77 | 0.11 | -0.17 |
| 3 | 0.30 | 0.18 | 0.69 | 0.09 | 0.52 |
| 4 | 0.45 | 0.04 | 0.70 | 0.10 | -0.04 |
| 5 | 0.41 | 0.90 | 0.67 | 0.12 | -0.18 |
| 6 | 0.40 | 0.90 | 0.77 | 0.12 | -0.17 |
| 7 | 0.40 | 0.24 | 0.80 | 0.08 | 0.52 |
| 8 | 0.50 | 0.05 | 0.76 | 0.08 | -0.01 |
| 9 | 0.97 | 0.59 | 0.45 | 0.08 | 0.54 |
| 10 | 1.27 | 0.76 | 0.66 | 0.09 | -0.67 |
| 11 | 0.79 | 0.41 | 0.66 | 0.03 | 0.51 |
| 12 | 1.52 | 0.04 | 0.80 | 0.08 | -0.01 |
| 13 | 1.98 | 0.49 | 0.42 | 0.01 | 0.38 |
| 14 | 2.00 | 1.00 | 0.33 | 0.03 | -0.94 |
| 15 | 1.52 | 0.04 | 0.80 | 0.08 | 0.00 |
| 16 | 1.77 | 0.28 | 0.80 | 0.10 | -0.71 |

Table 3. Optimal solution of the MBB example

| Component | $x_0$ | $y_0$ | $L/2$ | $t/2$ | $p$ |
|---|---|---|---|---|---|
| 1 | 0.49 | 0.94 | 0.76 | 0.12 | 0.01 |
| 2 | 0.53 | 0.94 | 0.80 | 0.11 | 0.00 |
| 3 | 0.59 | 0.08 | 0.76 | 0.13 | 0.00 |
| 4 | 0.53 | 0.08 | 0.72 | 0.13 | 0.01 |
| 5 | 0.43 | 0.98 | 0.69 | 0.09 | 0.04 |
| 6 | 0.31 | 0.79 | 0.71 | 0.06 | -0.62 |
| 7 | 0.52 | 0.08 | 0.68 | 0.13 | 0.01 |
| 8 | 0.70 | 0.03 | 0.76 | 0.09 | -0.01 |
| 9 | 0.46 | 0.94 | 0.71 | 0.11 | 0.00 |
| 10 | 1.00 | 0.93 | 0.70 | 0.11 | -0.01 |
| 11 | 0.76 | 0.02 | 0.80 | 0.08 | 0.02 |
| 12 | 0.56 | 0.08 | 0.74 | 0.13 | 0.00 |
| 13 | 1.45 | 0.58 | 0.61 | 0.05 | 0.76 |
| 14 | 1.91 | 0.81 | 0.63 | 0.08 | -0.55 |
| 15 | 1.68 | 0.04 | 0.50 | 0.08 | 0.01 |
| 16 | 1.10 | 0.18 | 0.55 | 0.09 | -0.52 |
| 17 | 2.32 | 0.67 | 0.36 | 0.01 | 0.92 |
| 18 | 2.15 | 0.64 | 0.58 | 0.08 | -0.56 |
| 19 | 2.12 | 0.04 | 0.61 | 0.09 | 0.03 |
| 20 | 1.66 | 0.04 | 0.48 | 0.08 | 0.01 |
| 21 | 2.98 | 0.71 | 0.34 | 0.01 | 0.66 |
| 22 | 3.00 | 0.80 | 0.34 | 0.01 | -0.69 |
| 23 | 2.94 | 0.16 | 0.35 | 0.04 | 0.61 |
| 24 | 2.78 | 0.20 | 0.59 | 0.10 | -0.60 |